Title: SLOPE INSTABILITY OF THE EARTHEN LEVEE IN BOSTON, UK: NUMERICAL SIMULATION AND SENSOR DATA ANALYSIS

Article Type: Research Paper

Corresponding Author: Mrs. Natalia Melnikova,

Corresponding Author's Institution: University of Amsterdam

First Author: Natalia Melnikova

Order of Authors: Natalia Melnikova; David Jordan; Valeria Krzhizhanovskaya, PhD, associated professor; Peter Sloot, PhD, Professor

Abstract: The paper presents a slope stability analysis for a heterogeneous earthen levee in Boston, UK, which is prone to occasional slope failures under tidal loads. Dynamic behavior of the levee under tidal fluctuations was simulated using a finite element model of variably saturated linear elastic perfectly plastic soil. Hydraulic conductivities of the soil strata have been calibrated according to piezometers readings, in order to obtain correct range of hydraulic loads in tidal mode. Finite element simulation was complemented with series of limit equilibrium analyses. Stability analyses have shown that slope failure occurs with the development of a circular slip surface located in the soft clay layer. Both models (FEM and LEM) confirm that the least stable hydraulic condition is the combination of the minimum river levels at low tide with the maximal saturation of soil layers. FEM results indicate that in winter time the levee is almost at its limit state, at the margin of safety (strength reduction factor values are 1.03 and 1.04 for the low-tide and high-tide phases, respectively); these results agree with real-life observations. The stability analyses have been implemented as real-time components integrated into the UrbanFlood early warning system for flood protection.

Dear Editor in Chief of Computers and Geotechnics,

Please find enclosed a manuscript titled " Slope Instability of the Earthen Levee in Boston, UK: Numerical Simulation and Sensor Data Analysis" for consideration.

We have developed a finite element module for stability analyses of earthen dikes which are monitored by an early warning system (EWS) for flood protection. The module (called *Virtual Dike*) employs live sensor data for models calibration and can be run with the real-time input from water level sensors or with predicted high water levels due to upcoming storm surge or river flood. The main objective of the module design is simulating "virtual" sensors time series for normal and abnormal modes (including different kinds of failure) and feeding the Artificial Intelligence component with the obtained series. In order to generate realistic data sets for the AI, the Virtual Dike module is extensively tested and validated on real-life dikes monitored by the early warning system (so far three dikes have been modelled: the Livedike in Groningen, the Netherlands; the Boston levee in UK and the experimental full-scale IJkDijk levee in Bad Nieuweschans, the Netherlands). In the proposed paper we present stability analysis of the Boston levee, which protects the town from floods caused by the raise of water in River Haven. Slope failures are occasionally observed at the site when tidal range reaches its maximum (at winter and spring times). The analysis results indicate that in winter conditions the dike is almost at its limit state, at the margin of safety. The results agree well with the real life observations.

We are convinced that this paper is of general interest to the readership of Computers and Geotechnics

Sincerely,

Natalia Melnikova on behalf of all co-authors.

# SLOPE INSTABILITY OF THE EARTHEN LEVEE IN BOSTON, UK: NUMERICAL SIMULATION AND SENSOR DATA ANALYSIS

N.B. Melnikova[a,b,c]*, D. Jordan[d], V.V. Krzhizhanovskaya[a,b,c], P.M.A. Sloot [a,b,e]

[a] *University of Amsterdam, The Netherlands*
[b] *National Research University ITMO, Russia*
[b] *Saint-Petersburg State Politechnical University, Russia*
[d] *HR Wallingford, United Kingdom*
[e] *NTU, Singapore*

The paper presents a slope stability analysis for a heterogeneous earthen levee in Boston, UK, which is prone to occasional slope failures under tidal loads. Dynamic behavior of the levee under tidal fluctuations was simulated using a finite element model of variably saturated linear elastic perfectly plastic soil. Hydraulic conductivities of the soil strata have been calibrated according to piezometers readings, in order to obtain correct range of hydraulic loads in tidal mode. Finite element simulation was complemented with series of limit equilibrium analyses. Stability analyses have shown that slope failure occurs with the development of a circular slip surface located in the soft clay layer. Both models (FEM and LEM) confirm that the least stable hydraulic condition is the combination of the minimum river levels at low tide with the maximal saturation of soil layers. FEM results indicate that in winter time the levee is almost at its limit state, at the margin of safety (strength reduction factor values are 1.03 and 1.04 for the low-tide and high-tide phases, respectively); these results agree with real-life observations. The stability analyses have been implemented as real-time components integrated into the *UrbanFlood* early warning system for flood protection.

## 1    Introduction

Recent catastrophic floods around the world have spawned a large number of projects aimed at the development of stronger and "smarter" flood protection systems. The design of Early Warning Systems (EWS) for flood protection and disaster management poses a grand challenge to scientific and engineering communities and involves the following fields of research:

- Sensor equipment design, installation and technical maintenance in flood defence systems;
- Information and Communication Technologies in application to:
  o gathering, processing and visualizing sensor data;
  o developing Common Information Space (CIS) middleware for connecting sensor data, relevant documents, analysis tools, modeling software and advanced scientific visualization;
  o providing Internet-based interactive access to CIS for researchers and maintenance personnel;
- Development of computational models and simulation components for stability analysis of flood protection barriers, failure probability evaluation, prediction of flood dynamics and ways for evacuation;
- Development of a decision support system for public authorities and citizens that will help making informed decisions in case of emergency and in routine levee quality assessment, thus reducing flood risk and providing advanced tools for flood management.

Many projects, among which are FLOODsite http://www.floodsite.net, Flood Control 2015 http://www.floodcontrol2015.com and development of the International Levee Handbook http://www.leveehandbook.net, attempt to solve some of the EWS aspects listed above. Our research was conducted under the UrbanFlood European 7[th] Framework Program project (http://www.urbanflood.eu), which unites the research on a physical study of levee failure mechanisms (Melnikova et al., 2013, 2011a, 2011b; Krzhizhanovskaya and Melnikova, 2012), general design and implementation of an EWS (Krzhizhanovskaya et al., 2011; Pengel et al., 2013), development of EWS components for simulation of dike breaching and flood spreading (Gouldby et al., 2010; Krzhizhanovskaya et al., 2013a,b), city evacuation (Mordvintsev et al., 2012), and development of an artificial intelligence (AI) system employing data-driven methods (Pyayt et al., 2013a,b) and its application in conjunction with the finite element analysis module (Pyayt et al., 2011; Erdbrink et al., 2012, 2013).

Finite element analysis of the Boston levee has been performed in a real-time component "*Virtual Dike*" of the *UrbanFlood* early warning system for flood protection (levee is also called "dike" in the Netherlands). The *Virtual Dike* component is a finite element module for numerical analyses of dikes macro-stability under hydraulic and mechanical loadings (Melnikova et al., 2013, 2011b). This component can be run with a real-time input from water level sensors or with predicted high water levels inferred from upcoming storm surge or river flood. In the first case, comparison of simulated pore pressures with real data can indicate a change in soil properties or in dike operational conditions (e.g. failure of a drainage facility). In the second case, simulation can predict the macro stability of the dike and indicate the "weak" spots in the dikes that require attention of dike managers and city authorities. Simulated dynamics of dike parameters (including local and overall stability, pore pressure, local stresses and displacements) describes the non-stationary behavior of the dike (changing over time). The results of the *Virtual Dike* simulations have been used to train an artificial intelligence system on "normal" and "abnormal" virtual sensor dynamics (Pyayt et al., 2011).

---

* Corresponding author. *E-mail address*: N.Melnikova@uva.nl, tel: +7(921)319-65-31

The final goal of the *UrbanFlood* project was the development of a flood early warning system that can monitor many kilometres of dikes. Therefore all computational modules had to be very efficient to meet the faster-than-real-time requirements. The resulting *Virtual Dike* software module for dike stability analysis is based on a linear elastic perfectly plastic Drucker-Prager soil model coupled with the Richards model for fluid flow in variably saturated porous media. For the Boston levee, these models predicted a correct shape of a slip surface (see section 5).

The conventional method for macro- stability analysis is the limit equilibrium method (LEM) although the FEM is increasingly used by designers/analysts. The first limit equilibrium method for slope stability analysis (Fellenius/Petterson method) was proposed in (Fellenius, 1927) and considers a balance of disturbing and stabilizing forces acting on vertical soil slices located above a circular slip surface. The ratio restoring forces/disturbing forces is termed the Factor of Safety. Later the method underwent numerous modifications, for instance by Taylor, Bishop, Morgenstern-Price, Spencer, Janbu and others (an extensive overview of existing LEM methods and their comparison can be found in (Fredlund and Krahn, 1977; Chen and Morgenstern, 1983; Duncan, 1996)). Some of those, like Fellenius' and Bishop's methods, assume a circular slip surface and do not satisfy horizontal equilibrium conditions for the soil slices, while Spencer's, Morgenstern-Price's and Janbu's generalized methods work with non-circular slip surfaces and satisfy all equilibrium conditions. In this paper we employ a Bishop's method (Bishop, 1955b) for slope stability analysis of the test levee located in Boston, UK (see section 2 for the test site description).

The finite element method was first applied to slope stability analysis in (Whitman and Bailey, 1967). Zienkiewicz (Zienkiewicz et al., 1975) has proposed to treat divergence of the numerical solution as the indicator of an actual slope failure, so that simulated failure occurs naturally as the shear strength of the soil becomes insufficient to resist the shear stresses. Zienkiewicz has also proposed a shear strength reduction method to evaluate the slope stability margin. In the shear strength reduction method, soil strength parameters are gradually scaled down until the onset of slope instability is reached - detected by the divergence of numerical analysis iterations. A strength reduction factor *SRF* is then defined as the ratio of the original and scaled strength parameters.

In FEM, no assumptions on the shape of the critical surface are made. Failure mechanisms occurring in heterogeneous soils (like for example simple shear in layered strata) can not be reproduced by limit equilibrium methods. For this particular case of the Boston levee, FEM simulation predicts an approximately circular slip surface (see section 5 below); hence, LEM application for the analysis of the case is reasonable, too.

A drawback of FEM method is that it requires knowledge of stiffness characteristics of soil (Young's modulus and Poisson's ratio), which are not always available. However, Griffiths and Lane have shown (Griffiths and Lane, 1999) that while stiffness characteristics affect displacements values, they do not significantly alter slope stability margin.

Levees subjected to tidal oscillations contain variably saturated zones. It is well known that pore suctions in vadose zones stabilize slopes (see, for example, Krahn et al., 1989; Griffiths and Lu, 2005). The extension of Terzaghi's classical effective stress principle on unsaturated soils was first proposed by Bishop (1955a) and then was extensively developed (e.g., Fredlund et al., 1978; Vanapalli et al., 1996). In these extensions, soil matric suction is relaxed by a matric suction coefficient (which is a function of soil-water saturation) and contributes to the calculation of effective stresses from total stresses. Sensitivity of slope safety margin to the pore suctions depends highly on the water table elevation, soil type and infiltration conditions (Griffiths and Lu, 2005). Case studies of unsaturated slope failures can be found in (Tsaparas et al., 2002; Cho and Lee, 2001). In the present work, we omitted pore suctions when calculating effective stresses above the phreatic line, assuming effective stresses to be equal to total stresses in the vadose zone. This assumption was based on the piezometers readings (see section 3 for details), which showed that the clayey part of the Boston levee volume stays fully saturated during tidal cycles, while a vadose zone exists on top of the dike in the loose, coarse-grained stratum built from the mixture of soil and debris and thus producing quite low suctions ("made ground" stratum in Figure 2).

In our research we have analyzed the slope stability of the Boston levee in the UK using LEM and FEM techniques. This heterogeneous earthen dike has a documented history of slope slippages under tidal load, which is up to 6m on spring tides. The Boston levee has been equipped with pore pressure and temperature sensors producing real-time output signals which are registered and stored in the Common Information Space (CIS) of the *UrbanFlood* early warning system for flood protection (Balis et al., 2011). The finite element analysis of porous flow and deformations in the dike has been performed within the EWS workflow in the *Virtual Dike* module. A limit equilibrium analysis of slope stability has been performed for a number of different tidal phases in a separate computational module employing *Geo-Stability* software.

The paper is organized as follows: section 2 contains a description of the test-site, ground conditions and the levee construction. Sensors and their measurements are described in section 3; section 4 describes mathematical models involved in the modeling and numerical implementation details. Simulation results and cross-validation of the two methods are presented in section 5, followed by conclusions (section 6).

**2  Test site description and ground conditions**

The earthen levee at Boston, a town on the east coast of England at high risk of flooding, is known for a history of frequent toe slippages on the river side. This mechanism is presumably caused by high pore water pressures remaining in the dike when the river water recedes. The dike forms the right bank of the River Haven, which has a tidal range of about 6m. The crest level of the dike is at 6m above the mean sea level; the deepest part of the river bed is at 2m from the mean sea level. The site is predominately grassed with several trees (Figure 1a). It has suffered from instability at the toe along the majority of its length (Figure 1b).

The area was investigated in 2010 as part of the Boston Barrier Phase 1 Ground Investigation, this included a single borehole within the study dike. Further boreholes and Cone Penetration Tests were carried out as part of the installation of the sensors for the UrbanFlood Project. From the investigation, the variation in ground conditions across the site may be summarized as shown in Figure 2; beneath made ground and a thin layer of fine sand, lies some 5m of soft to firm alluvial clays. These in turn overlie sands and stiff boulder clay. Sensor locations are specified in Figure 2 relative to the Ordnance Datum (OD), which is the reference sea level in Great Britain (defined as the mean sea level at Newlyn in Cornwall between 1915 and 1921).

Regrettably, few samples were recovered from the sensor installation boreholes and none were tested. Accordingly as part of the early stages of analyses, back-analyses using LEM were carried out with various strengths for the soft clays in order to replicate the toe failures observed in the dike. The most plausible values were friction angle  =25° and cohesion *c*=2 kPa, typical acceptable values for soft clay (Verruijt, 2001); these were used for subsequent calculations. For stiff alluvial clay, sand and made ground typical reference values have been chosen (see Table 1).

Table 1. Summary of soil parameters for the Boston levee

| Property | Made ground | Fine sand | Soft brown clay | Dark brown sand | Firm grey clay |
|---|---|---|---|---|---|
| Hydraulic conductivity, m/day | 10 | 1 | 0.05 | 10 | 0.01 |
| Van Genuchten parameter α, 1/m | 2 | 2 | 0.5 | -* | - |
| Van Genuchten parameter n | 1.5 | 1.5 | 1.5 | | |
| Saturated water fraction | 0.4 | 0.4 | 0.43 | | |
| Residual water fraction | 0.045 | 0.045 | 0.3 | | |
| Density dry kN/m3 | 19.5 | 17.5 | - | | |
| Density wet kN/m3 | - | | | 18 | 22 | 18 |
| Young's modulus, MPa - drained | 18 | 28 | 2 | 18 | 2 |
| Young's modulus, MPa - undrained | - | - | 2.2 | - | 2.3 |
| Poisson's ratio - drained | 0.35 | 0.3 | 0.35 | 0.3 | 0.3 |
| Poisson's ratio - undrained | - | - | 0.49 | - | 0.49 |
| Cohesion, kPa | 5 | 0 | 2 | 0 | 5 |
| Friction angle, grad | 30 | 28 | 25 | 27.6 | 23.5 |

*Table cells containing "-" refer to the properties not used in corresponding soil strata. For example, Van Genuchten parameters for dark brown sand and firm grey clay were not used in simulation because these soil strata stay saturated during tide oscillations.

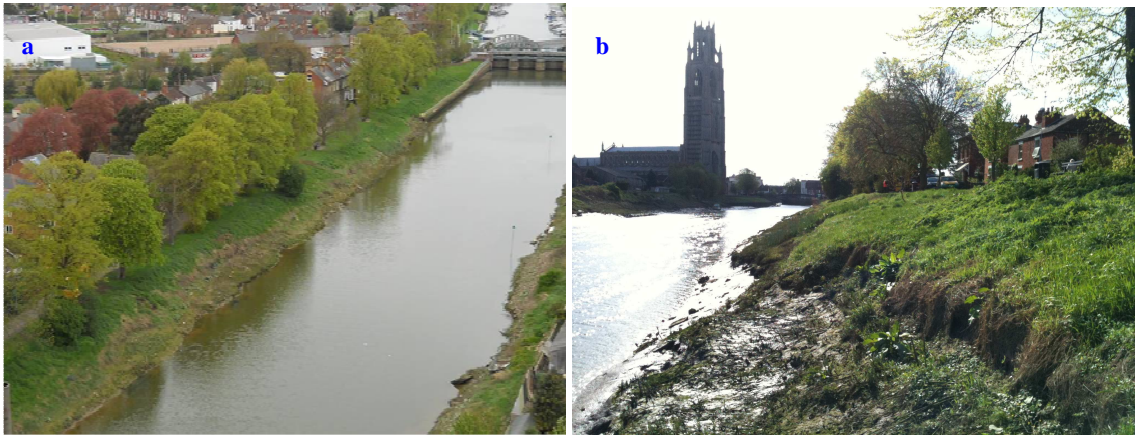

Figure 1. (a) Boston levee location at the right bank of the river Haven; (b) View of embankment with signs of toe slippage

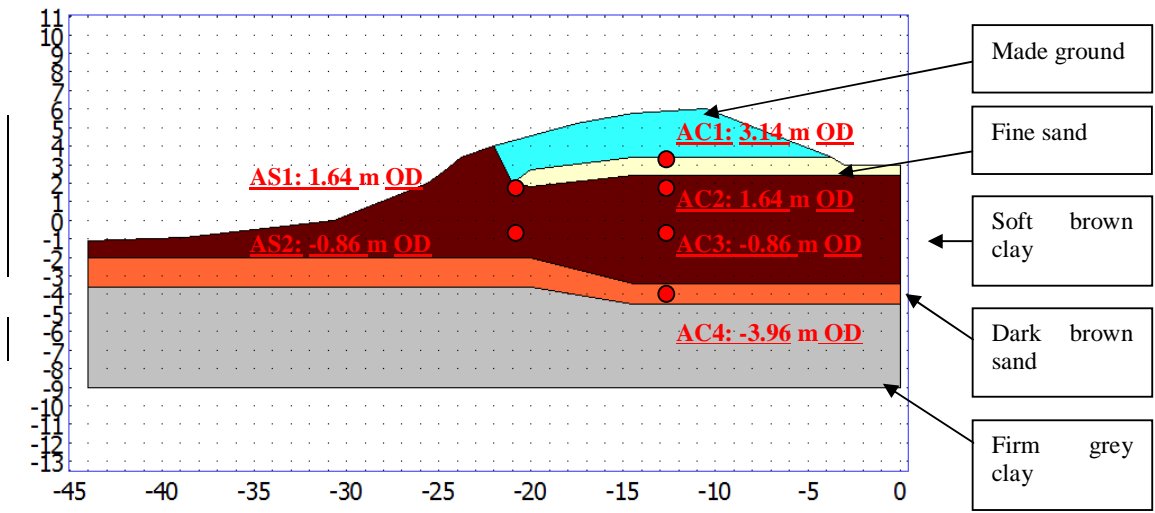

Figure 2. Scheme of the cross-section A: soil build-up and sensors elevations, metres from Ordnance Datum

## 3   Instrumentation and sensor data analysis

The dike has been equipped at each of two cross-sections with seven GeoBeads MEMS (micro-electro-mechanical) sensors registering pore pressure and media temperature. Differences in temperature measurement curves strongly indicate water flow through the soil - any drop in sensor temperature might be an indication of the development of piping. The Geobeads sensors have been installed in boreholes in the two planar transversal cross-sections (depicted as A and B in Figure 3). Positions of the GeoBeads sensors in the central cross-section A are shown in Figure 2; vertical elevations are specified relative to the mean sea level and to the local ground level. Sensors AC1-AC4 are of increasing depth at the crest of the bund whilst sensors AS1-AS2 are located at about mid-slope height.

The instrumentation and control building for the Grand Sluice on Haven River provided an ideal location for situating the computer equipment and providing power to the sensors. The use of mains power eliminated the need to replace batteries and reduced the level of maintenance required on the sensor equipment, ensuring uninterrupted data signals. The control building is located at 129 m distance upstream from cross-section A and is shown in the top of Figure 3 near the bridge with the sluice.

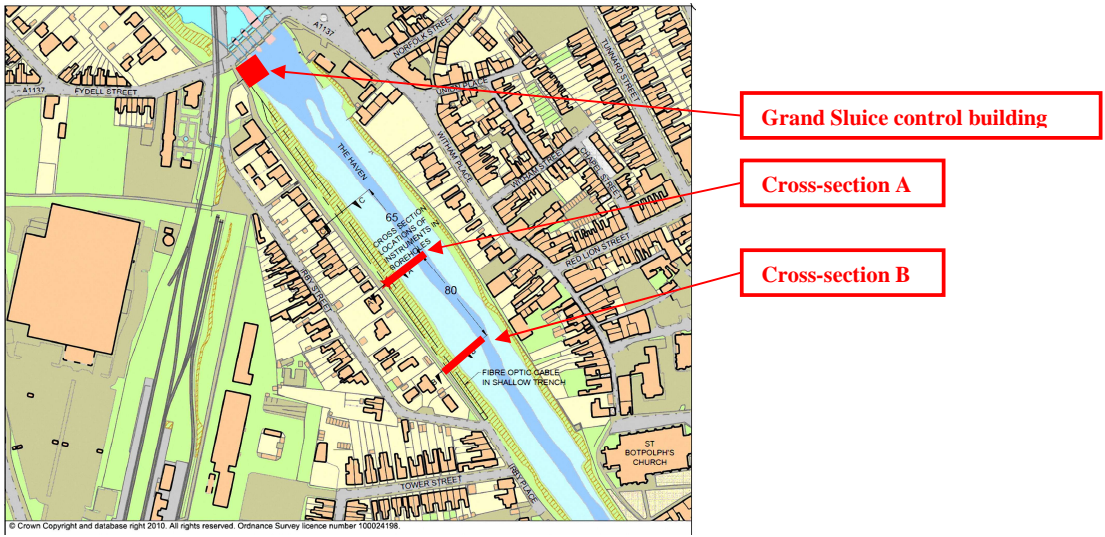

Figure 3. Boston levee site, with the cross-section locations

Sensor readings in the Boston levee had been previously analyzed in (Simm et al., 2012), for a period of one year since the installation of sensors in May 2011. In the present research we focus on the response of the dike to tidal loading (putting aside slow seasonal processes), so we analyzed sensor data and dike stability for a one-month period in January 2012. The readings in cross-section A are presented in Figure 4. Figure 4a shows the water level dynamics in the river. Figure 4b gives the pore pressure dynamics inside the dike.

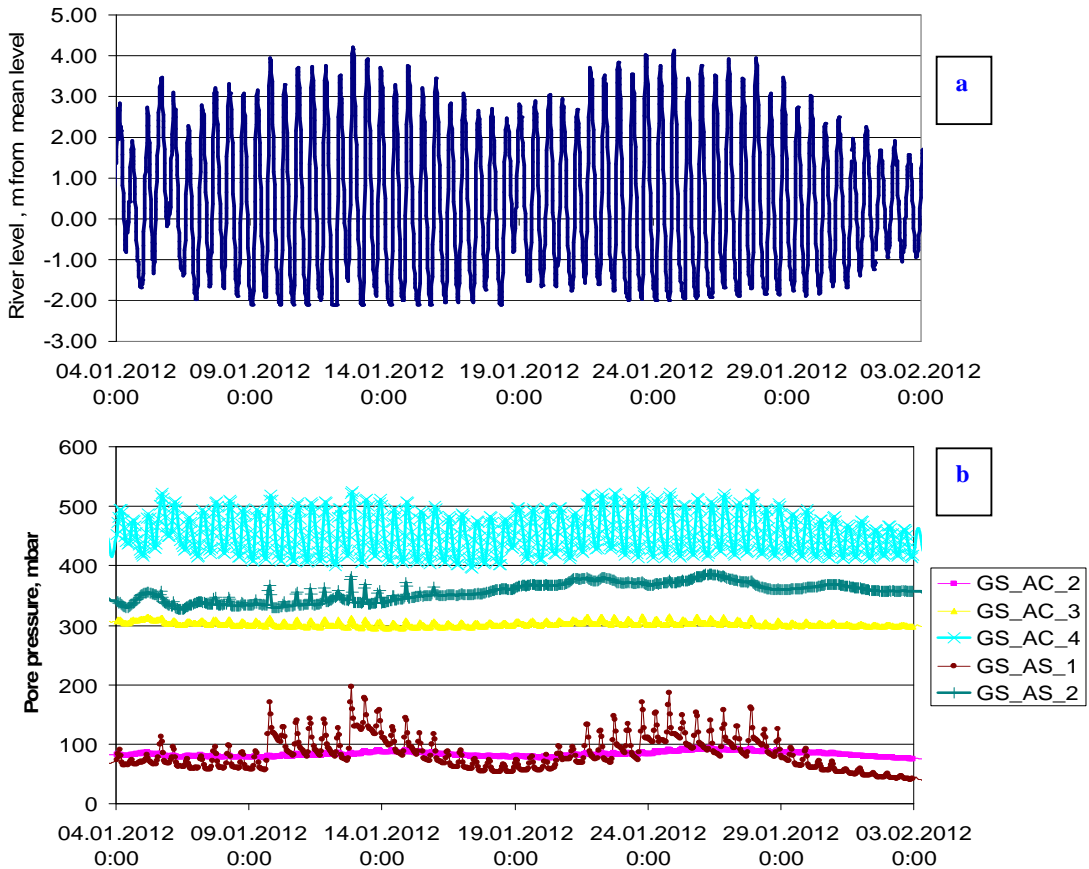

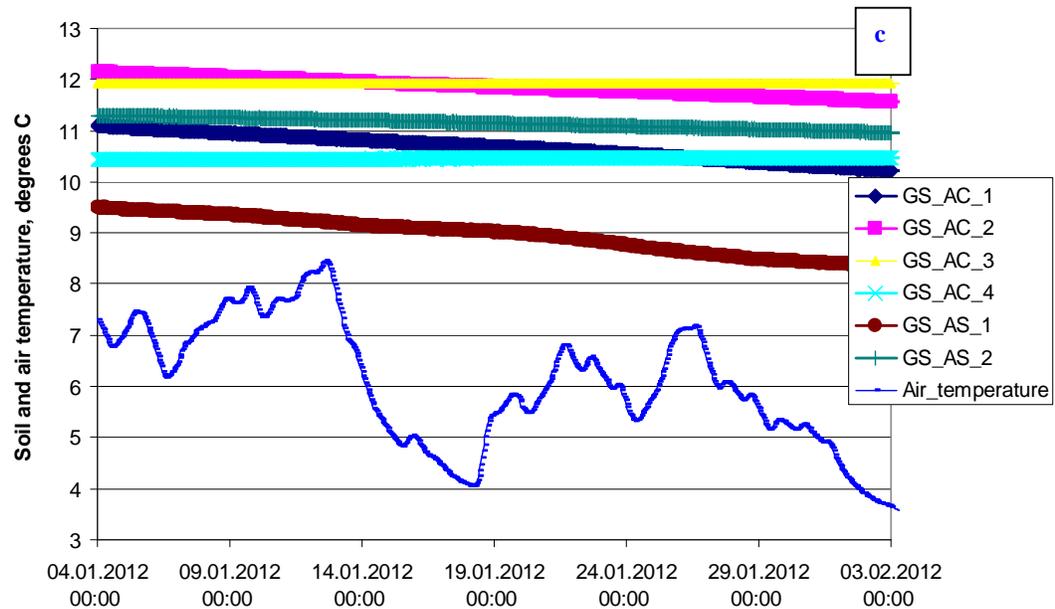

Figure 4 (a) river level dynamics; (b, c) sensor readings: (b) pore pressure; (c) temperature

The results show a good response to tidal variations particularly in AS1, the upper sensor on the slope. Plateau-like segments of the AS1 curve correspond to the low-tide phases: the large soft brown clay layer stays saturated even at a low-tide phase (when the river bed is almost dry) - hence the pressure distribution and hydraulic load in the clay layer do not change during the low-tide phase. There is no significant time-lag between the tide and piezometric levels. The tidal range in mid-river is about 6m whilst the AS1 pressure head range is about 1m (100 mbar). This limited magnitude of response reflects the position of the piezometers within the slope relative to the maximum tidal variation in the centre of the river channel. It is considered that the piezometers are measuring an undrained elastic response in the soil-water continuum due to the loading caused by variation in water levels in the adjacent river channel. The assumption on the undrained state of the clay stratum is confirmed by comparing mean pressure heads measured by sensors in the soft brown clay stratum with pressure heads in the river and in the underlying dark brown sand layer (the comparison is discussed below).

According to sensor readings, AC1 is located in a dry zone above the ground water level, which is why it is not shown in Figure 4b. AC2 and AC3 sensors located in the soft clay layer far from the river-side slope are almost insensitive to the tide. Mean value of pore pressure oscillations (relative to atmospheric pressure) in sensors AC2 and AS1 is about 90 mbar. Both sensors are located at the same elevation of 1.64 metres above OD – so the mean ground water table in the dike is at 2.54 m above OD. According to the soil build-up scheme (Figure 2), the phreatic line goes through the thin layer of fine sand.

AC4 naturally produces high response to the tidal loading due to high conductivity of the dark brown sand layer where the sensor is placed. Mean pore pressure in AC4 is 450 mbar and its elevation is 3.96 m below OD, which gives mean value of total head +0.54 m above OD. This value agrees with the mean river level (Figure 4b), which is 0.6 for the considered period of time, proving that the dark brown sand layer is hydraulically connected to the river. As it was mentioned above, in the upper layers of soft brown clay, fine sand and made ground, the mean total head is at 2.54 m above OD which is two meter higher than in the underlying sand layer. This discrepancy confirms the assumption that the clay layer and the overlying soil layers (fine sand, made ground) are hydraulically isolated both from the foundation sand layer and from the river due to low permeability of clay. The massive clay layer retains large amount of water coming with precipitation which is most intensive during winter time.

Soil and air temperature curves measured by sensors are shown in Figure 4c. In winter time the soil temperature is higher than the air temperature due to the high heat capacity of soils. Sensor AS1 produces minimal temperature values as it is closer to the land surface than the other sensors. AC1-AC2 and AS1-AS2 temperatures gradually decrease during the month, together with the dynamics of the mean air temperature value (and presumably of the water temperature in the river). Sensors AC3-AC5 located in deep soil strata below the ground water level produce nearly constant values of temperature around 10-12°C.

In the case of piping erosion in the dike, temperature sensors can provide warning information: decrease of local temperature value from expected soil temperature to the water temperature indicates that piping is occurring in the dike (Pyayt et al., 2014). However piping has never been visually observed or sensor detected at this site. As it was mentioned above, the dike is prone to occasional river-side toe slippages, such as shown in Figure 1b. There is no data available from the maintenance records about the precise moments in time when these local slippages occurred.

A more detailed analysis of sensor data has been reported in (Pyayt et al., 2013a), where a data-driven approach with a neural network were applied for modelling a transfer function between the sensors within the Artificial Intelligence software module.

## 4 Mathematical models and numerical implementation

For the analysis of the Boston levee stability, two approaches have been employed: finite element modeling and limit equilibrium method. Below we separately describe mathematical models and numerical solution procedures for the two approaches.

### 4.1 Finite element model

In FEM analysis, a one-way coupled fluid-to-structure interaction model for the planar cross-section of the dike has been considered.

Water flow through the porous media is described by Richards' equation (Bear, 1979) with the van Genuchten model for water retention in vadose zones (Van Genuchten, 1980):

$$(C + \theta_e S)\frac{\partial p}{\partial t} + \nabla \cdot [-\frac{K_S}{\mu} k_r \nabla (p + \rho g z)] = 0 \tag{1}$$

where $C$, $\theta_e$, $S$ are specific moisture capacity, effective water content and specific storage, respectively; storage coefficient $S$ is computed as the inverse of the soil skeleton bulk modulus $K$: $S=1/K$; $p$ is water pressure (negative in unsaturated zone); t is time; $K_S$ is permeability of saturated media; $k_r=k_r(p)$ is relative permeability; $\mu$ is dynamic viscosity of water; $g$, $\rho$, $z$ are standard gravity, water density and vertical elevation coordinate, respectively.

Specific moisture capacity $C$ and relative permeability $k_r$ in the unsaturated zone are defined as functions of the effective water content (Van Genuchten, 1980).

The mechanical sub-model describes stress-strain state of the dike under hydraulic load, gravity and volumetric pore pressure load obtained from flow simulation. Linear elastic perfectly plastic strains of the soil skeleton are described by the general equations of plastic flow theory (Hill, 1950):

$$\begin{cases} \nabla \cdot \underline{\underline{\sigma}} + \rho_s \underline{g} = 0 \\ \underline{\underline{\sigma}} = \underline{\underline{\sigma}}_{eff} - p \underline{\underline{I}} \\ \underline{\underline{\sigma}}_{eff} = \frac{E}{1+\nu}\left[\frac{\nu}{(1-2\nu)}\varepsilon \underline{\underline{I}} + \underline{\underline{\varepsilon}}\right], & \text{if } F < 0, \\ \underline{\underline{\dot{\varepsilon}}}_{pl} = -q\frac{\partial F}{\partial \underline{\underline{\sigma}}}, \ \underline{\underline{\dot{\varepsilon}}} = \underline{\underline{\dot{\varepsilon}}}_{pl} + \frac{1}{3K}\dot{I}_1 \underline{\underline{I}}, & \text{if } F = 0 \end{cases} \tag{2}$$

where $\nabla = \underline{e}_x \frac{\partial}{\partial x} + \underline{e}_y \frac{\partial}{\partial y} + \underline{e}_z \frac{\partial}{\partial z}$ is gradient operator; $\rho_s$ is soil density; $\underline{g}$ is gravity vector; $\underline{\underline{\sigma}}$ and $\underline{\underline{\sigma}}_{eff}$ are total and effective stress tensors, respectively (compressive stresses are negative); $E$ is Young's modulus; $\nu$ is Poisson's ratio; $\varepsilon = \varepsilon_{xx} + \varepsilon_{yy} + \varepsilon_{zz}$ is volume deformation (positive for expansion); $\underline{\underline{I}}$ is unit tensor; $\underline{\underline{\varepsilon}} = (\nabla \underline{U} + (\nabla \underline{U}^T)/2$ is deformation tensor; $\underline{U}$ is vector of displacements; $\underline{\underline{\dot{\varepsilon}}}_{pl}$ is plastic deformation rate tensor; q is plastic multiplier; $F$ is plastic yield function; $K = \frac{E}{3(1-2\nu)}$ is bulk modulus; $I_1 = \sigma_{xx\,eff} + \sigma_{yy\,eff} + \sigma_{zz\,eff}$ is the first effective stress invariant.

Plastic flow has been modeled with a modification of the Drucker-Prager plasticity model, optimized for plane strain problems by providing the best smooth approximation of the Mohr-Coulomb surface in the stress space (Chen and Mizuno, 1990):

$$F = \alpha \cdot I_1 + \sqrt{J_2} - F_{DP},$$

where $J_2 = I_1^2/3 - I_2$ is second deviatoric effective stress invariant, $I_2 = \sigma_{xx\,eff} \cdot \sigma_{yy\,eff} + \sigma_{zz\,eff} \cdot \sigma_{yy\,eff} + \sigma_{xx\,eff} \cdot \sigma_{zz\,eff} - \sigma_{xy\,eff}^2$ is second effective stress invariant; $\alpha$ and $F_{DP}$ are constants: $\alpha = tg(\varphi)/\sqrt{9 + 12 \cdot tg^2(\varphi)}$, $F_{DP} = 3c/\sqrt{9 + 12 \cdot tg^2(\varphi)}$; $c, \varphi$ are cohesion and internal friction angle, respectively.

Equations (1, 2) form a one-way coupled flow-structure interaction system, where the porous flow sub-model generates a volume load (computed as pore pressure gradient) for the mechanical sub-model.

Boundary conditions for fluid and mechanical sub-models are schematically shown in Figure 5a,b.

Hydraulic boundary conditions are listed below:

- Black line (Figure 5a) - the river side, pressure boundary condition: $\begin{cases} p = \rho g \cdot (h(t) - y) & \text{for } y \leq h(t), \\ p = 0 & \text{for } y > h(t) \end{cases}$,

    where $h(t)$ is river level, metres;

- Cyan line (Figure 5a) - the land side, pressure boundary condition: $\begin{cases} p = \rho g \cdot (h_L(t) - y) & \text{for } y \leq h_L(t), \\ p = 0 & \text{for } y > h_L(t) \end{cases}$, where $h_L(t)$ is ground water level at the land side.

Ground water table at the land side was specified using mean values of AC2 and AC4 pore pressure readings discussed in the previous section. For the dark brown sand and firm grey clay layers, $h_L = 0.5$ m; for the soft brown clay layer and overlying layers, $h_L = 2.5$ m.

Magenta line shows impervious walls.

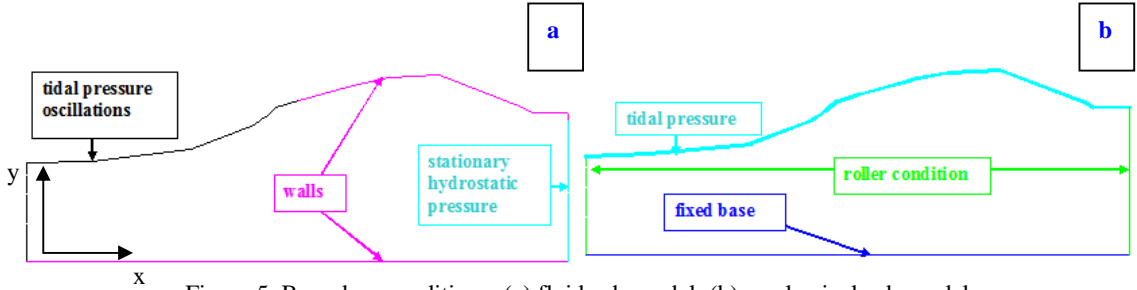

Figure 5. Boundary conditions; (a) fluid sub-model; (b) mechanical sub-model

Mechanical boundary conditions have been specified as follows:

- Cyan line (Figure 5b) - free surface with total normal stress specified: $\{n \cdot \underline{\underline{\sigma}} = -\rho g \cdot (h(t) - y) \text{ for } y \leq h(t); \; n \cdot \underline{\underline{\sigma}} = 0 \text{ for } y > h(t)\}$;

- Green line - roller condition: normal displacements and shear stresses are zero $U_x = 0$, $\sigma_{xy} = 0$;

- Blue line - fixed condition: zero displacements $\underline{U} = 0$.

The overall solution procedure includes two loading stages:

1) Gravity settlement problem solution with stationary hydraulic boundary conditions: both river level and land side ground water level are stationary: $h_L = 0.5$ m (dark brown sand and firm grey clay layers); $h_L = 2.5$ m (soft brown clay layer and the overlying layers). Initial condition: above the phreatic line - $p = -5$kPa; below the phreatic line - hydrostatic pore pressure distribution according to water tables specified in the boundary conditions. Gravity load and buoyancy load were applied incrementally to an initially stress-free domain. Mechanical behavior of clay layers (brown soft clay and firm grey clay) was simulated as "drained" at this stage. Stresses obtained at this stage were used in the next stage to define pre-stresses in the domain.

2) Tidal mode simulation. Initial condition: pore pressures and stresses are obtained from the previous completed stage; displacements are zero. At each physical time step, a filtration problem was solved with a time-dependent FE solver; than the obtained pore pressures were passed to the mechanical sub-model as volume load. Increments of hydraulic loads were computed and gradually applied in the incremental parametric solver. Mechanical behavior of clay layers was simulated as undrained at this stage. The undrained properties of clay layers were assigned as follows: the Poisson's ratio $\nu_u = 0.49$; the Young's modulus $E_u = E(1+\nu_u)/(1+\nu)$, where $\nu$, $E$ are drained properties. For each tide phase, strength reduction factors *SRF* were calculated by scaling all the cohesions and friction angles' tangencies in all the domains until the critical state (slope instability) was achieved under a fixed load:

$$SRF = c/c_{m\arg in} = \tan\varphi / \tan\varphi_{m\arg in},$$

where $c_{m\arg in}, \varphi_{m\arg in}$ are cohesion and friction angle at the onset of instability, under original, not scaled loads.

Computations have been carried out using finite element software package Comsol. The functionality of this package supports multi-physics coupling – equations describing the time-dependent fluid sub-model and the quasi-static mechanical sub-model can be united in one system of equations, which will be integrated by a time-dependent solver. However, convergence of the time-dependent solver was poor for the highly-nonlinear mechanical sub-model: integration in time-domain required very small time-steps and finally stopped due to divergence of non-linear iterations. Typically plastic deformation problems in Comsol require a parametric solver for their solution. That's why a two-step solution scheme was used, where the porous flow sub-model was solved by a time-dependent solver; then pore pressure loads were transferred to the mechanical sub-model which in turn was solved by parametric solver. The computational workflow is described in Figure 6: the workflow with calls to Comsol routines is processed in a Matlab script; data exchange between the sub-models occurs in the computer RAM memory within the Matlab run-time environment. The loop in Figure 6 corresponds to time-stepping; stagger iterations were not used for the solution process, as the problem is one-way coupled.

A two-dimensional finite element mesh was composed of about 15000 triangular elements, and the second order of space approximation was used for all loading stages.

The FEM-analysis *Virtual Dike* module has been deployed on a cloud computing resources of the SURFsara BiG Grid High Performance Computing and e-Science Support Center https://www.surfsara.nl/. The cloud is hosted on a 128-core cluster and uses OpenNebula open source cloud computing management toolkit with KVM Virtual machine software. Simulations are run under Ubuntu Linux in a shared memory parallel mode. The parallel performance results have been published in (Melnikova et al., 2011a).

### 4.2 Limit equilibrium model

The *Geo-Stability* software package has been used for the limit equilibrium analysis. This uses the well known Bishop method of slices (Bishop, 1955b) to calculate the factor of safety of the dike's slope. The factor of safety (*FoS*), describing the capacity of a slope to withstand its own weight together with applied external loadings from surcharge and groundwater, is defined as the ratio of restoring forces $F_R$ (soil shear strength + externally applied restoring forces) to disturbing forces $F_D$ (soil self weight + externally applied disturbing forces): $FoS = F_R / F_D$, where the sums of restoring and disturbing forces are calculated on all possible circular slip surfaces with arbitrary diameters and centre locations. The critical failure surface corresponds to that providing the minimal value of *FoS*.

By assuming varying groundwater profiles within the dike for varying external water conditions, the module was able to provide a matrix of factors of safety for varying water levels (see Table 3 in the next section). This matrix of results relates external water levels and pore pressures (as measured in the sensors) with factor of safety. Additional values of factor of safety are then determined by an interpolation routine that finds intermediate values between those within the look-up table. Development of such stability matrices or look-up tables that could be interrogated using actual sensor values significantly simplified the process of using sensor information compared to the approach originally envisaged. Hence, the look-up table approach has been adopted within the study.

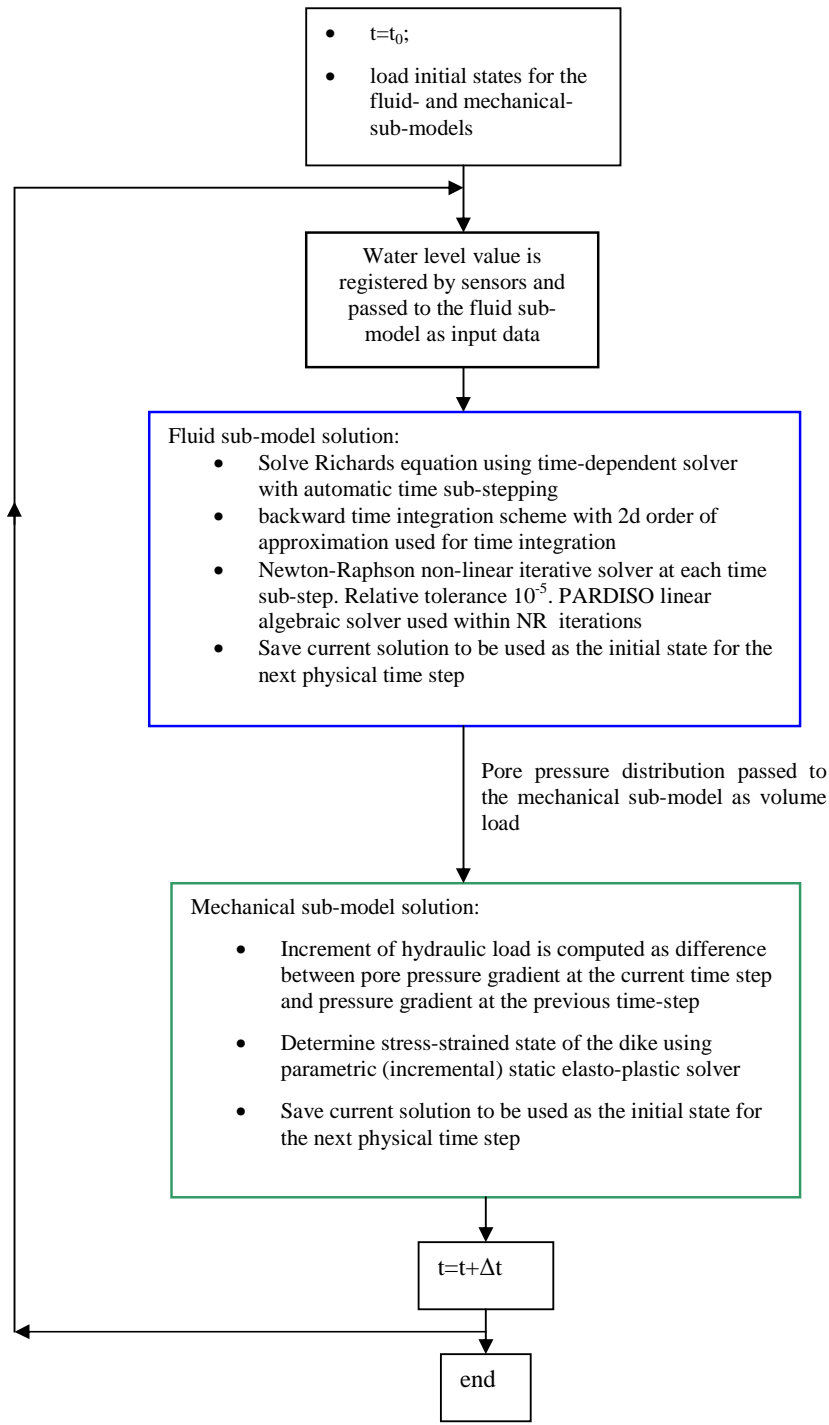

Figure 6. Tidal mode simulation: computational workflow in the FE package

## 5 Simulation results

### 5.1 Finite element simulation results

First, the *Virtual Dike* FEM model has been extensively tested and calibrated. Hydraulic conductivity of the dark brown sand layer has been adjusted by matching pore pressure dynamics measured by real sensors with simulated dynamics. The calibration procedure has been described in (Melnikova et al,, 2013). The river level dynamics during the simulation period is presented in Figure 7a. A comparison of real and simulated signals is presented in Figure 7b for the AC4 sensor located in a dark brown sand layer. The calibrated value of conductivity is given in Table 1.

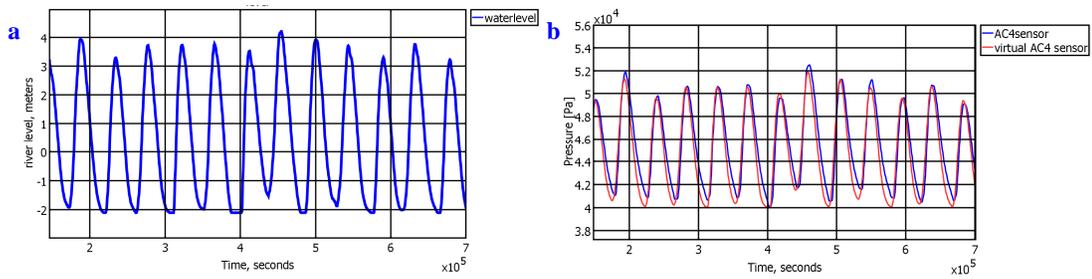

Figure 7. River level dynamics during simulation period (a); comparison of real and simulated signals is presented (b),

The calibrated *Virtual Dike* FEM model then was used for stability analysis. Below we present the simulation results. Figure 8a shows pore pressure distributions at high tide and low tide. In the clay layer it is only a local zone at the river slope that changes pore pressure distribution with tidal oscillations. Maximal response to the tide is naturally observed in the base sand layer. Effective saturation distribution changes very little with the tide and looks similar for the low tide and high tide phases (Figure 8c) - the thick layer of soft clay located between the levels of -2 m and +4 m does not really change saturation during relatively fast diurnal oscillations, hence the hydraulic load in clay does not change much from high tide to low tide.

Effective plastic strain distributions in the deformed domain are presented in Figure 8b for high tide and low tide phases. Simulations have converged, indicating that the dike is stable under the tidal load. However, formation of a shallow slip surface located entirely in the soft clay layer can be easily seen at the low tide (Figure 8c).

Total displacements distributions are shown in Figure 8d. At high tide, total displacements are mostly produced by vertical flotation of the dike; at low tide, the slip surface formation is clearly identified at the river slope.

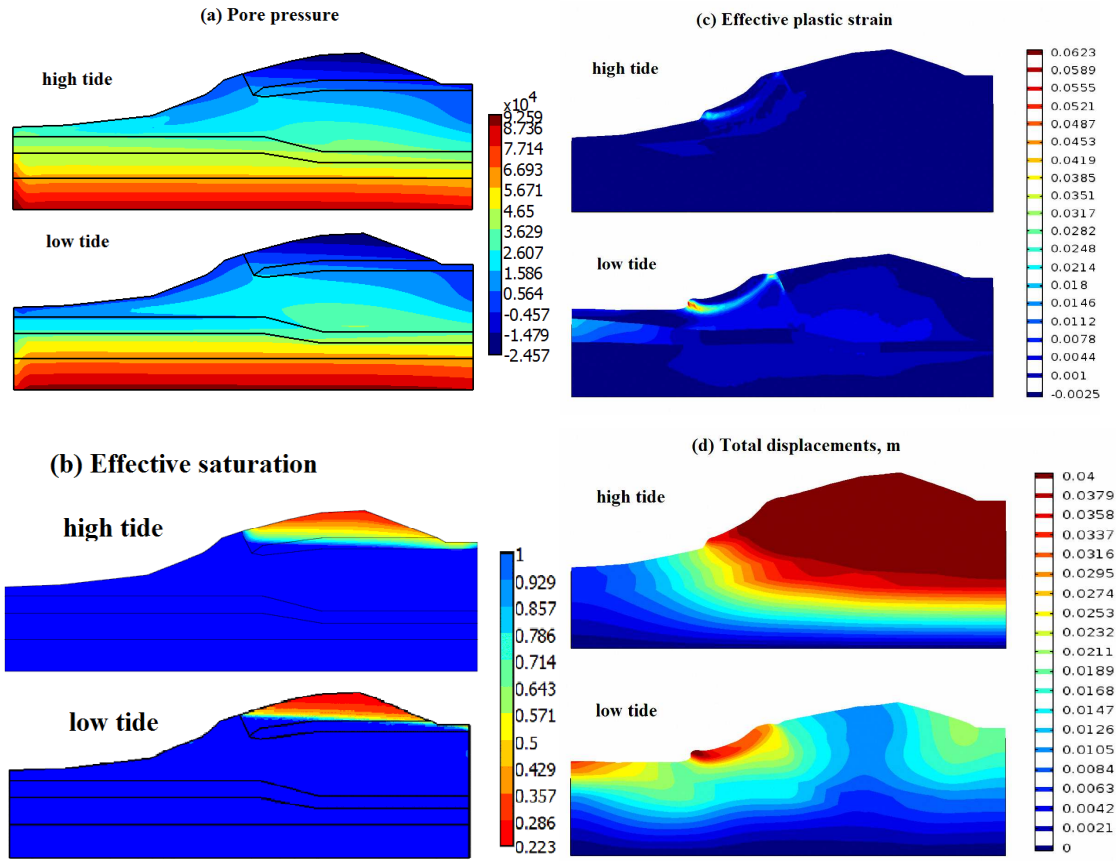

Figure 8. (a) Pore pressure distribution; (b) Effective saturation; (c) Effective plastic strains (displacements scaling factor 50); (d) Total displacements: (displacements scaling factor 50).

## 5.2 Limit equilibrium modelling results

Safety factors obtained in the *Geo-Stability* LEM analysis are presented in Table 2, in the form of a look-up table with variable river levels (RL) and ground water levels (GWL) within the dike. Columns AC1-AS2 show pore pressure values calculated at sensor locations. In the limit equilibrium analysis, the phreatic line shape was determined as a straight line connecting river and land side water levels. This simplification caused a significant difference in the distribution of pore pressures and soil weights in the dike, compared with the FEM modeling results, as the clay layer is less saturated in LEM for the most of the GWL conditions.

Figure 9c shows that the FEM is predicting at low tide an approximately circular failure surface indicating that use of the LEM is appropriate, though because of the simplifications required modeling for the latter different results were obtained from the two methods – see Section 5.3.

The lowest value of the Factor of Safety *FoS* =1.04 has been obtained for the combination of the low tide condition RL = 0m with the highest ground water levels (GWL = 4m and GWL = 6m). This fact confirms the conclusion made from the FEM simulation results presented in the previous section: the less stable mode corresponds to the combination of maximal saturation of the dike with the low-tide water levels.

Slip surfaces (circles) obtained by Bishop's method for the high tide phase (RL=4 m, GWL=0 m) and for the low tide phase (RL=-1.1 m, GWL=0 m) are shown in Figure 9b,d. The slip surfaces are shallow, with the high-tide slip circle having much larger radius than the low-tide circle. The shape of the low-tide slip surface agrees well with the real-life slippage observations.

Table 2. Stability factor values for various hydraulic conditions

| | | Sensor measurements in cross-section A | | | | | | | |
|---|---|---|---|---|---|---|---|---|---|
| River level (RL) | Assumed ground water level (GWL) | AC1 | AC2 | AC3 | AC4 | AC5 | AS1 | AS2 | Safety factor |
| [m] | [m] | [kPa] | [kPa] | [kPa] | [kPa] | [kPa] | [kPa] | [kPa] | [-] |
| -1.1 | 0 | Atmos | Atmos | 8 | 39 | 63 | Atmos | 7.94 | 1.55 |
| 0 | 0 | Atmos | Atmos | 8 | 39 | 63 | Atmos | 7.94 | 1.515 |
| 0 | 2 | Atmos | 3 | 28 | 59 | 83 | 2.94 | 27.94 | 1.28 |
| 0 | 4 | 8 | 23 | 48 | 79 | 103 | 22 | 47 | 1.04 |
| 0 | 6 | 28 | 43 | 68 | 99 | 123 | 22 | 47 | 1.04 |
| 2 | 0 | Atmos | Atmos | 8 | 39 | 63 | Atmos | 7.94 | 1.67 |
| 2 | 2 | Atmos | 3 | 28 | 59 | 83 | 2.94 | 27.94 | 1.55 |
| 2 | 4 | 8 | 23 | 48 | 79 | 103 | 22 | 47 | 1.08 |
| 2 | 6 | 28 | 43 | 68 | 99 | 123 | 22 | 47 | 1.08 |
| 4 | 0 | Atmos | Atmos | 8 | 39 | 63 | Atmos | 7.94 | 2.11 |
| 4 | 2 | Atmos | 3 | 28 | 59 | 83 | 2.94 | 27.94 | 2.11 |
| 4 | 4 | 8 | 23 | 48 | 79 | 103 | 22.94 | 47.94 | 1.88 |
| 4 | 6 | 28 | 43 | 68 | 99 | 123 | 22.94 | 47.94 | 1.6 |

## 5.3 Comparison of FEM and LEM results

Comparison of the results obtained by the two analysis techniques for the low-tide and high-tide phases is presented in Table 3. In *Geo-Stability* LEM analysis, the lowest ground water level (GWL) was assumed at 0 m. In *Virtual Dike* FEM, GWL was slightly different being equal to 0.4 m. River level (RL) during high tide was +4 m above mean sea level; at low tide, RL was -1.1 m. For the cross-validation of LEM and FEM models, factors of safety for GWL=0.4 m and RL=-1.1 m have been obtained by interpolating between values in Table 2.

Table 3. Safety factors calculated by *Virtual Dike* FEM and *Geo-Stability* LEM.

| High tide (RL=4 m, GWL=0.4 m) | | Low tide (RL=-1.1 m, GWL=0.4 m) | |
|---|---|---|---|
| *Virtual Dike* (FEM) SRF | *Geo-Stability* (LEM) FoS | *Virtual Dike* (FEM) SRF | *Geo-Stability* (LEM) FoS |
| 1.04 | 2.11 | 1.03 | 1.55 |

The values of *FoS* obtained in *Geo-Stability* LEM program are much higher (by 50% at low tide and by 100% at high tide) than the values of strength reduction factors *SRF* obtained by the *Virtual Dike* FEM.

Values of *SRF* obtained by FEM do not significantly differ for the high tide (RL=4 m) and for the low tide (RL=-1.1 m). This is due to the fact that a thick layer of soft clay located between the levels of -2 m and +4 mOD at the river slope has low permeability and does not significantly change saturation during relatively fast diurnal oscillations (see Figure 8b). We believe that in reality the hydraulic load changes quite insignificantly with the tide due to the large amount of clay in the dike (this is confirmed by sensor readings in Figure 4b) and this effect has been reproduced in the more realistic FEM simulation. In the case of continuous rainfall infiltration, the stability factors will most likely decrease due to saturation of the top made ground layer and the increase in the weight of the dike, which can result in slope slippage.

In *Geo-Stability* LEM analysis hydraulic loads differ significantly for the high tide and low tide, as the ground water table was assumed to be a straight line connecting river and landside water levels. Due to the same reason, the difference between high and low tide critical slip surfaces in *Geo-Stability* (Figure 9b,d) is much higher than in the *Virtual Dike* (Figure 9a,c). In both models the critical slip surface is shallow. For the high tide, it ends slightly above the river side toe of the dike, while for the low tide, the radius of the surface increases and slip surface ends below the river side toe. The radiuses of slip surfaces increase from high tide to low tide, both in *Virtual Dike* and in *Geo-Stability* programs. However, in the *Virtual Dike* FEM model the critical surfaces are located entirely within the soft clay layer whilst in the *Geo-Stability* LEM model they cross the fine sand and made-ground layers.

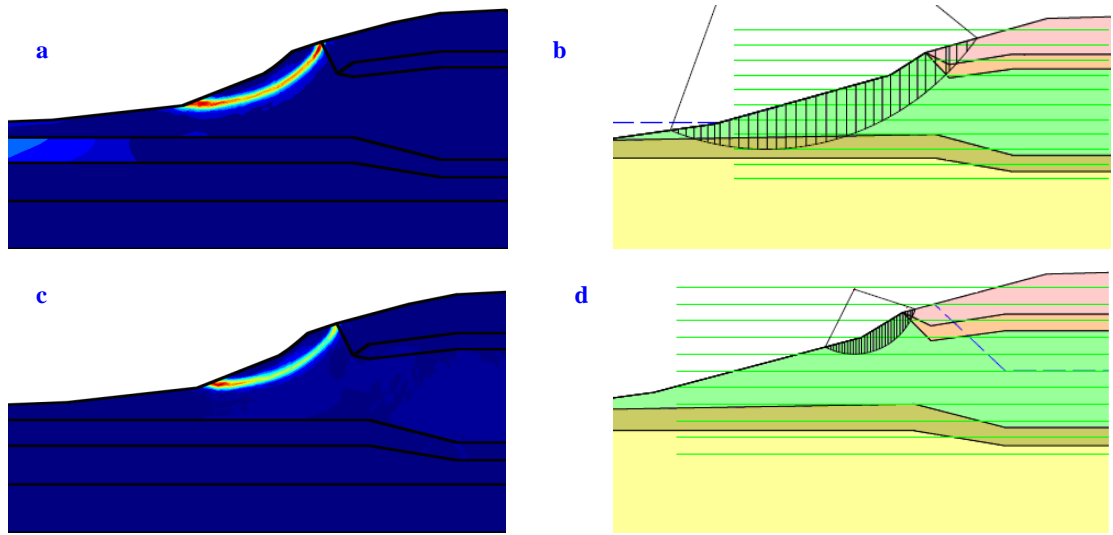

Figure 9. Slip surfaces in critical state; high tide: (a) FEM, (b) LEM; low tide: (c) FEM, (d) LEM

## 6 Conclusions

Two computational modules for the analysis of earthen levees stability have been developed and integrated into the *UrbanFlood* early warning system for flood protection. The modules employ finite element modeling and limit equilibrium analysis, respectively. Simulations of dike stability can be run with the real-time input from water level sensors or with predicted high water levels due to upcoming storm surge or river flood. In the first case, comparison of simulated pore pressures with real data can indicate a change in soil properties or in dike operational conditions (e.g. failure of a drainage facility). In the second case, simulation can predict the structural stability of the dike and indicate the "weak" spots in the dikes that require attention of dike managers and city authorities.

The two modules have been validated on a number of test sites, including the Boston levee site monitored by the early warning system in real-time mode. The Boston levee protects the town from floods caused by the raise of water in River Haven. Slope failures are occasionally observed at the site when tidal range reaches its maximum (at winter and spring times).

Hydraulic conductivities of the dike layers have been calibrated according to piezometers readings, in order to obtain correct range of hydraulic loads. The readings also show that the massive soft brown clay layer stays fully saturated during tidal cycles and its condition is undrained. According to the stability analyses carried by finite element method and by Bishop's limit equilibrium method, slope failure occurs with the development of an approximately circular slip surface located in the soft brown clay layer. Both models, LEM and FEM, confirm that the least stable hydraulic condition is the combination of the minimum river levels at low tide with the maximum saturation of soil layers. The factors of safety calculated by Bishop's limit equilibrium method are significantly higher than strength reduction factors calculated by FEM (by 50-100 %). In our case, the discrepancy between LEM and FEM results is predominantly due to the differences in calculation of hydraulic loads in the dike from tidal oscillations. FEM takes into account water storage in clay layers at low tide phase, thus the hydraulic load is determined more accurately than in LEM. FEM results indicate that in real-life winter conditions the dike is almost at its limit state, at the margin of safety (strength reduction factor values are 1.03 and 1.04 for the low-tide and high-tide phases, respectively).

**Acknowledgements**

This work was supported by the EU FP7 project *UrbanFlood*, grant N 248767; by Government of Russian Federation: contract 11.G34.31.0019 under Leading Scientist Programme and Grant 074-U01 under "5-100-2020" Programme; by the BiG Grid project BG-020-10, # 2010/01550/NCF with financial support from the Netherlands Organization for Scientific Research NWO. It was carried out in collaboration with AlertSolutions, Siemens, TNO, Deltares, SURFsara Computing and Networking Services.